**Spin-charge-entangled non-Fermi liquid in a candidate material for a doped spin liquid**

H. Oike, [1]*† Y. Suzuki, [1] H. Taniguchi, [2] K. Miyagawa, [1] K. Kanoda[1]*

[1]Department of Applied Physics, University of Tokyo, Bunkyo-ku, Tokyo 113-0032, Japan.

[2]Department of Physics, Saitama University, Saitama, Saitama 338-8570, Japan.

*Corresponding author. E-mail: hiroshi.oike@riken.jp (H.O.); kanoda@ap.t.u-tokyo.ac.jp (K.K.)

†Current addresses: RIKEN Center for Emergent Matter Science (CEMS), Wako, Saitama 351-0198, Japan.

**Abstract**: Quantum spin liquids are exotic Mott insulators that carry extraordinary spin excitations and thus, when doped, expected to afford novel metallic states coupled to the unconventional magnetic excitations. The organic triangular-lattice system $\kappa$-(ET)$_4$Hg$_{2.89}$Br$_8$ is a promising candidate for the doped spin-liquid and hosts a non-Fermi liquid at low pressures. We show that, in the non-Fermi liquid regime, the charge transport confined in the layer gets deconfined sharply at low temperatures, coinciding with the entrance of spins into a quantum regime as signified by a steep decrease in spin susceptibility behaving like the triangular-lattice Heisenberg model indicative of spin-charge separation at high temperatures. This suggests a new type of non-Fermi liquid, where interlayer charge-deconfinement is associated with spin-charge entanglement.

**Main Text**

Strongly interacting electrons show a variety of ground states and distinctive elementary excitations, depending on the strength of interactions, band filling, and lattice geometry. In half-filled systems, on-site Coulomb interaction $U$, when exceeding bandwidth $W$, prohibits electrons from doubly-occupying a site, and drives the system into a Mott insulating state [1]. Mott insulators typically exhibit magnetic orders at low temperatures. However, several materials with quasi-triangular lattices host spin liquids without magnetic ordering due to spin frustration [2-8] and exotic spin states are theoretically suggested [9-16]. Carrier injection into Mott insulators by doping gives rise to metallic and superconducting states with unconventional properties [17-24]. To our knowledge, however, there is no experimental challenges of doping spin liquids, although a doped spin liquid is proposed as a model for high-$T_c$ superconductivity in copper oxides, which are however antiferromagnets before being doped. Thus, what phases emerge when spin liquids are doped is an open issue.

A promising candidate of doped spin liquids is the organic conductor $\kappa$-(ET)$_4$Hg$_{2.89}$Br$_8$ [abbreviated as $\kappa$-HgBr] with a triangular lattice, where ET denotes bis(ethylenedithio)-tetrathiafulvalene [25]. As depicted in Fig. 1(a), $\kappa$-HgBr has conducting ET layers sandwiched by insulating layers composed of Hg and Br ions. In the conducting layers, ET dimers form a nearly isotropic triangular lattice [26-29] with a $t'/t$ value close to unity as in the spin liquid material $\kappa$-(ET)$_2$Cu$_2$(CN)$_3$ (abbreviated as $\kappa$-Cu$_2$(CN)$_3$ hereafter), where $t$ and $t'$ are transfer integrals between antibonding orbitals of dimers, as indicated in Fig. 1(a), which form a conduction band [29]. The band filling deviates from a half by the deficiency of the Hg composition from 3.0 [30]; that is, 11% of holes are introduced to the half-filled band. In-plane resistivity $\rho_{//}$ shows a non-Fermi liquid behavior at low pressures. Owing to the highly compressible nature of organic materials, $U/W$ is varied in a wide range by pressure. As pressure increases at low temperatures, the non-Fermi liquid turns into a Fermi Liquid (FL) at approximately 0.5 GPa [31, 32], where the Hall coefficient shows a distinct change indicative of an increase in effective carrier density in spite of the fixed band filling [32]. Noticeably, $\rho_{//}$ keeps increasing with temperature over the Ioffe-Regel limit of several mΩcm even in the high pressure regime, indicating that the so-called strange metal is extended in a wide pressure range at high temperatures as has been observed in many strongly correlated metals [33]. These transport properties suggest a transition or crossover from a doped Mott insulator with double occupancy strongly prohibited to a correlated metal with all electrons contributing to carriers (Fig. 1(b)), in analogy to the Mott metal-insulator transition at half filling [32]. Thus, $\kappa$-HgBr is a doped spin-liquid candidate with tunable electron correlation.

In the present study, we investigate out-of-plane transport, which measures the inter-layer tunneling of quasiparticles. For example, out-of-plane transport captured a variation in charge excitations from quasiparticle nature to collective one in layered cobalt oxides [34], and the quasi-particle density of states in the zero-mode Landau level in layered Dirac-electron systems [35], while the interpretation of the in-plane transport is not straightforward. In underdoped copper oxides, for another example, pseudogap formation is captures by non-metallic behavior in optical conductivity and resistivity in the out-of-plane direction while in-plane resistivity shows metallic behavior [36]. This behavior indicates that quasiparticle excitations require a finite gap possibly arising from spin-charge separation into spinless holons and chargeless spinons, which have to be recombined for out-of-plane conduction [19]. Thus, out-of-plane transport probes the nature of the electron fluid in the plane. We measured the out-of-plane resistivity of $\kappa$-HgBr with tuning pressure across the non-FL-to FL transition/crossover point and profiled the degree of the out-of-plane metallicity in the temperature-pressure plane. We also measured spin susceptibility to know the spin states. It is revealed that how to acquire the out-of-plane metallicity is different between the non-FL regime and the FL regime. In the latter regime, charge transport is metallic in every direction, as expected in conventional Fermi liquids. Contrastingly, in the former non-FL regime, metallic transport is confined in the plane; however, it becomes deconfined abruptly at low temperatures, coinciding with the entrance of the frustrated spins into a quantum regime. A novel type of spin-charge-recoupled liquid is suggested.

The single crystals of $\kappa$-HgBr used in the present study were grown by standard electrochemical methods. Magnetic susceptibility was measured using a SQUID magnetometer (Quantum Design MPMS XL-7). The spin susceptibility is obtained by subtracting the core-diamagnetic contribution from the measured susceptibility. In-plane and out-of-plane resistivities were measured by conventional four probe method. Pressure was applied with the use of a dual structured clamp-type cell formed by BeCu and NiCrAl cylinders. Daphne7373 oil was used as a pressure-transmitting medium.

Figure 2 shows the temperature dependence of spin susceptibility $\chi_\text{spin}$, which is featured by temperature-linear dependence at high temperatures and a rounded peak at 30-40 K, followed by a sharp decrease, as reported in refs. [37, 38]. The present measurements have found the negligible anisotropy, which reflects non-significant spin-orbit interactions common to organic materials. The behavior of the $\chi_\text{spin}$ considerably differs from that of isostructural metallic compounds with half-filled bands, $\kappa$-(ET)$_2$X (X=Cu(NCS)$_2$ and Cu[N(CN)$_2$]Br); they show only weakly temperature-dependent Pauli-paramagnetic susceptibility of the order of $4\times10^{-4}$ emu/mol, which is less than a half of the peak value ($9\times10^{-4}$ emu/mol) in $\kappa$-HgBr, above 100 K [39, 40]. The temperature dependence of the $\chi_\text{spin}$ is well reproduced by the series expansion of the triangular-lattice Heisenberg model with the Pade approximation of order [7/7] [41], as shown in Fig.2. The nearest-neighbor exchange interaction, $J$, is approximately 140 K. The fact that the spin degrees of freedom behave like Heisenberg spins despite the metallic state strongly suggests spin-charge separation in a doped Mott insulator. Noticeably, the $\chi_\text{spin}$ nearly follows the behavior of the quantum spin-liquid Mott-insulator $\kappa$-Cu$_2$(CN)$_3$, which is also well reproduced by the triangular-lattice Heisenberg model with $J$ of 250 K [2]; the difference in $J$ is reasonably explained by differences in $U$ and $t$ between $\kappa$-HgBr and $\kappa$-Cu$_2$(CN)$_3$, the $t^{\,2}/U$ values of which are 4.6 meV and 6.7 meV, respectively, according to band calculations based on the extended Huckel method and tight-binding approximations with the dimer model [28, 29]. Thus, $\kappa$-HgBr is reasonably assumed to host a doped spin liquid.

Figure 3 shows the in-plane and out-of-plane resistivities. At ambient pressure, this material shows a number of spurious jumps in resistivity on temperature variation very probably due to the micro cracking in the crystal, as encountered in several organic materials. Thus, applying finite pressures was necessary for obtaining reliable data. At 0.1 GPa and 0.15 GPa, the temperature dependence of out-of-plane resistivity $\rho_\perp$ is non-metallic in a wide temperature range and takes a peak structure at approximately 20 K, followed by a steep decrease. Above 0.2 GPa, the peak is broadened and shifted toward higher temperatures. At pressures above 0.5GPa, the temperature dependence of $\rho_\perp$ is metallic in a wide temperature range and the metallicity increases with pressure. Considering that the $\rho_{//}$ behaves metallic in the whole pressure-temperature range studied, it is remarkable that there exist a wide temperature region in which only $\rho_\perp$ is non-metallic in the non-FL regime below 0.5 GPa.

To characterize the out-of-plane metallicity, we take the logarithmic derivative of $\rho_\perp$ with respect to temperature, $d(\ln\rho_\perp)/dT$. Figure 4(a) shows the pressure dependence of $d(\ln\rho_\perp)/dT$ at fixed temperatures. As pressure is decreased, the interlayer metallicity turns to decrease at approximately 0.5 GPa for every temperature except below 20 K. Figure 4(b) shows the contour plot of $d(\ln\rho_\perp)/dT$, which profiles the interlayer metallicity in the temperature-pressure plane. It is obvious that the interlayer-metallic region is extended to higher temperatures for higher pressures. To characterize the temperature or energy scale of the interlayer metallicity, we tentatively define $T_m$ as a temperature below which $d(\ln\rho_\perp)/dT$ exceeds 0.04. The pressure dependence of Tm shown in Fig. 4(b) suggests that there are three distinctive regions; $P < 0.3$ GPa, $0.3 < P < 0.6$ GPa and $P > 0.6$ GPa. The high-pressure region ($P > 0.6$ GPa) corresponds to the FL region identified by the behavior of $\rho_{//}$ and actually $T_m$ roughly coincides with the temperature where $\rho_{//}$ starts to depart from the FL behavior [32]. The intermediate region ($0.3 < P < 0.6$ GPa), in which $T_m$ steeply decreases with pressure, is a transient region to the regime of the doped Mott-insulator with non-FL character, in which Tm levels off to a low but finite value. It is peculiar that the nonmetal-to-metal crossover in the out-of-plane direction occurs sharply at low temperatures, 10-15 K (Fig.4(b)) in the non-FL regime.

To examine the possible involvement of superconducting fluctuations in this anomaly, we measured $\rho_{//}$ and $\rho_\perp$ with applying magnetic fields perpendicular to the layers (see Fig.5). It is obvious that, above $T_c$, both of $\rho_{//}$ and $\rho_\perp$ are not affected by the magnetic fields that are large enough to extinguish the superconductivity, indicating that conventional superconducting fluctuations are not pertinent to the low-temperature restoration of the interlayer coupling. On top of that, it is remarkable that the non-FL nature persists down to the lowest temperature studied, 1.8 K, when superconductivity disappears, contrary to the FL behaviors of non-doped systems in the vicinity of Mott transition [42].

A clue to the puzzling interlayer transport at low pressures is found in the behavior of $\chi_{spin}$ which exhibits a steep decrease in the same temperature range as $\rho_\perp$ does (see Fig.2). The similar decrease in $\chi_{spin}$ is observed in the spin-liquid system, $\kappa$-(ET)$_2$Cu$_2$(CN)$_3$, as well and argued to signify the entrance of the interacting spins into a quantum regime, in which a Fermi-degenerate system of spinons is argued to be a possible state [10]. If such a situation occurs in the spin sector in the present system, spinons get hybridized with doped holons in the quantum regime at low temperatures while they are separated at high temperatures. At low temperatures, there may be a case in which the spinons and holons form composite particles that can tunnel across the layers; however, they should not be the conventional quasiparticles because FL is not stabilized. Thus, the non-FL with charge excitation deconfined from a layer may be a novel quantum fluid that emerges from doping a spin liquid. Alternatively, it is not ruled out that the positive d(ln$\rho_\perp$)/d$T$ reflects the process, in which the hybridization of the spinons and holons develop into conventional quasiparticles on cooling. If this is the case, the ground state would be a FL with an extremely small Fermi energy-an extraordinary situation. Anyway, the low-temperature behaviors of $\rho_\perp$ and $\rho_{//}$ of the present doped triangular lattice are not straightforwardly understood in the conventional frame, thus appearing to invoke a novel notion.

The present results await to be theoretically investigated in the light of doped spin liquids. There are theoretical suggestions that exotic metallic phases can emerge in multi-band systems with conduction electrons and localized spins on the geometrically frustrated lattices [43-46] such as the Shastry-Sutherland lattice [47, 48] and pyrochlore lattices [49]. When the Kondo coupling between conduction electrons and localized spins are strong, they form a FL of a heavy mass [44-46]. As the coupling is decreased, however, electrons in the conduction band get only weakly hybridized with the frustrated spins and can coexist with spin liquids [44-46]. In some cases, the conduction electrons are predicted to partially contribute to Fermi volume, resulting in a fractionalized FL accompanied by $S = 1/2$ deconfined spinon excitations [43]. The present single-band system, in which only HOMO orbitals of ET molecules contribute to both transport and magnetic properties, is distinctive from this model, but show intriguing similarities; the anomalous metal at low pressures and at low temperatures emerges from a situation, in which the conducting and magnetic channels appear separated at high energies, and turns into a conventional FL when transfer integrals, which correspond to the Kondo coupling [50], is increased by applying pressure. In the present case, a single band is in charge of both of electrical conduction and magnetism. Whether and how the fractionalization in the multiband model is reduced to the single band may be a way to address the present issue.

The notion that mobile electrons, even if strongly interacting with each other, behave as a Fermi liquid is a basis for understanding metals. To seek for qualitatively different metallic "phases" from the Fermi liquid is a fundamental challenge to condensed matter physics. The Tomonaga-Luttinger liquid in one dimension is a known example, and other possibilities have been sought after in various circumstances, under which electrons reside [51-54]. The present results suggest that doped triangular lattices offer an unprecedented realm to bring about a novel non-Fermi liquid phase in a single band; namely, an interlayer-deconfined non-Fermi liquid with possible spinon-holon entanglement. Notably, superconductivity emerges in this situation. How Cooper pairing occurs in this strange metal is a forthcoming issue.

**Acknowledgements** We thank T. Koretsune, T. Senthil, N. Nagaosa, M. Ogata and J. Ibuka for useful comments. This work was supported in part by JSPS KAKENHI under Grant Nos. 20110002, 25220709, 24654101, and 11J09324 and the US National Science Foundation under Grant No. PHYS-1066293 and the hospitality of the Aspen Center for Physics.

**Figure legends**

**Figure 1** Crystal structure and schematic band filling-$U/W$ diagram. (a) Crystal structure of $\kappa$-$(ET)_4Hg_{2.89}Br_8$ viewed along c axis. Lower panel shows the conducting ET layer, which is parallel to b-c plane. The ET layers are sandwiched by insulating layers composed of Br and Hg ions. The Hg ions form chains along c axis, which is incommensurate with a sublattice comprised of Br ions and ET molecules, resulting in the non-stoichiometric composition, which is precisely known by the incommensurability determined by x-ray diffraction but cannot be varied because the non-stoichiometry is determined by the solid state chemistry between ET, Br and Hg. The electronic bands of $\kappa$-ET compounds are well described by the tight binding of antibonding molecular orbitals in the ET dimers (indicated by broken circles). The lattice of dimers is modeled to an anisotropic triangular lattice formed by transfer integrals, $t$ and $t'$ as depicted in the lower panel [29]. The anisotropy of the triangular lattice, $t'/t$, of the present system is estimated at 1.02 [26-29]. (b) Schematic band filling-$U/W$ phase diagram. The undoped Mott insulator, $\kappa$-$(ET)_2Cu_2(CN)_3$, and the doped one, $\kappa$-$(ET)_4Hg_{2.89}Br_8$, are located by green and red arrows, respectively, under pressure variation, while High-$T_c$ cuprates under doping are located by a black arrow.

**Figure 2** Temperature dependence of spin susceptibility of $\kappa$-(ET)$_4$Hg$_{2.89}$Br$_8$. The susceptibility under magnetic fields parallel and perpendicular to the conducting layers is indicated by red and black points. The broken lines represent the numerical curves obtained by the series expansion of the triangular-lattice Heisenberg model with Pade approximant of order [7/7] [41]. The comparison of the calculations and the experimental data yields the $J$ value of 130-150 K. Inset shows the temperature dependence of the spin susceptibility below 100 K.

**Figure 3** In-plane and out-of-plane resistivities of $\kappa$-(ET)$_4$Hg$_{2.89}$Br$_8$. (a and b) Temperature dependences of in-plane resistivity $\rho_{//}$ (a) and out-of-plane resistivity $\rho_{\perp}$ (b) at several pressures.

**Figure 4** Pressure- and temperature-dependences of interlayer metallicity. (a) Pressure dependence of dln$\rho_{\perp}$/d$T$. (b) Contour plot of dln$\rho_{\perp}$/d$T$ in the pressure-temperature plane. The broken line represents temperatures where dln$\rho_{\perp}$/d$T$ is equal to 0.04. The solid line represents superconducting transition temperatures determined by ac susceptibility measurements [32].

**Figure 5** Magnetic field-dependence of in-plane and out-of-plane resistivities of $\kappa$-

(ET)$_4$Hg$_{2.89}$Br$_8$ at low temperatures in the non-Fermi liquid regime. (a and b) In-plane resistivity $\rho_{//}$ (a) and out-of-plane resistivity $\rho_\perp$ (b) under magnetic fields of 0, 3, 6 and 9 T applied perpendicular to the conducting plane at a pressure of 0.3 GPa.

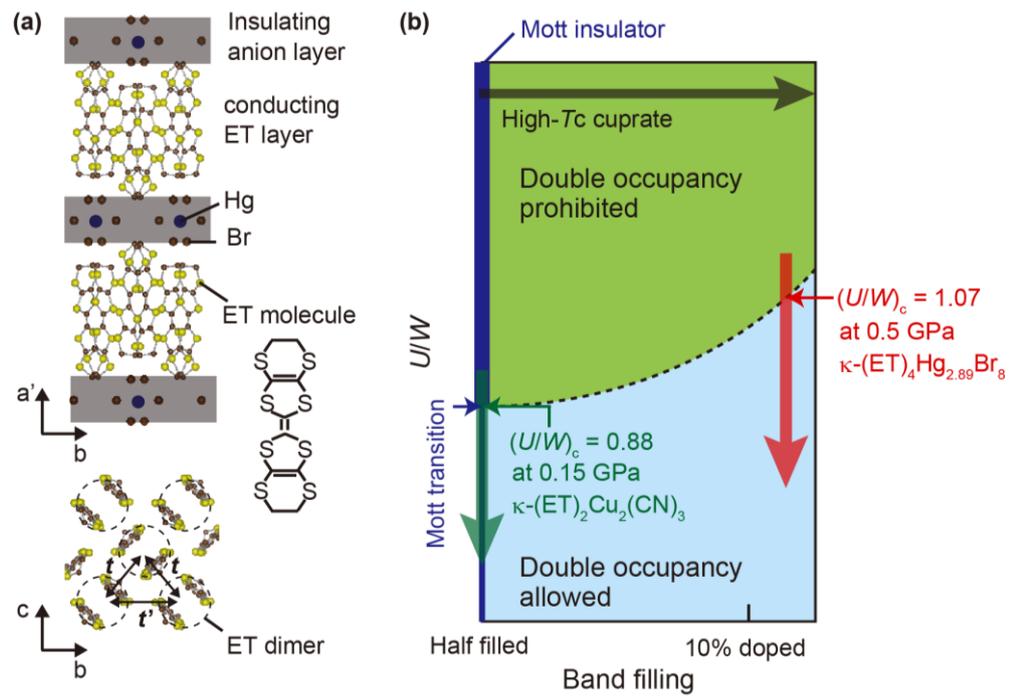

Fig. 1 Oike et al.

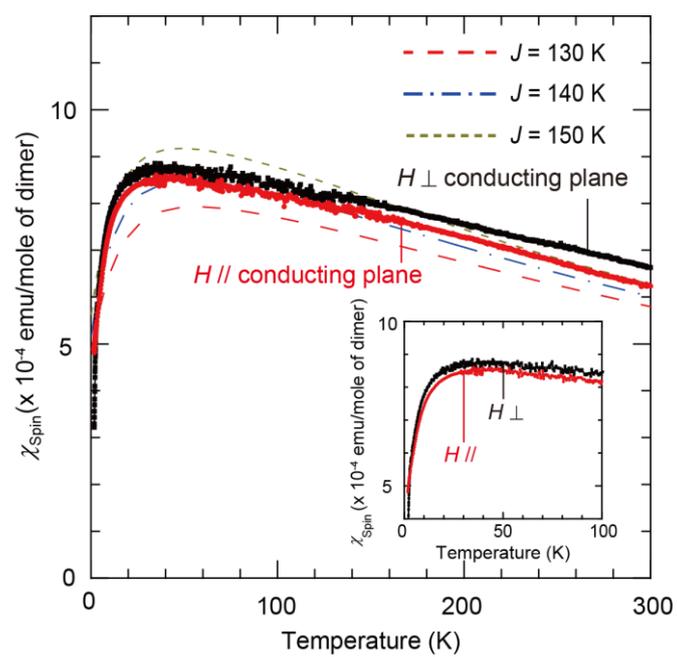

Fig. 2 Oike et al.

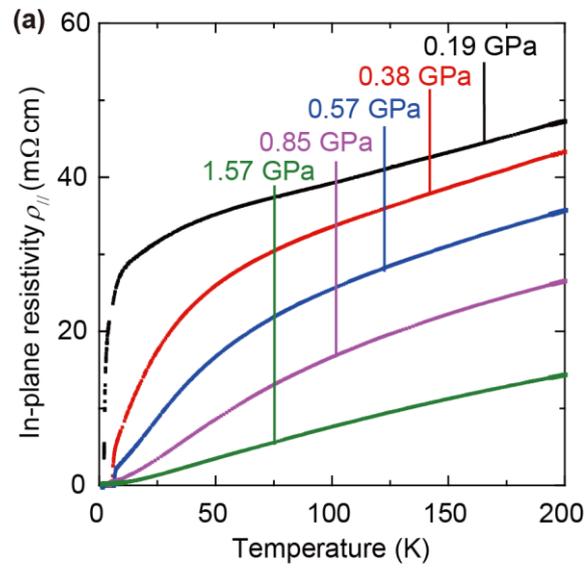

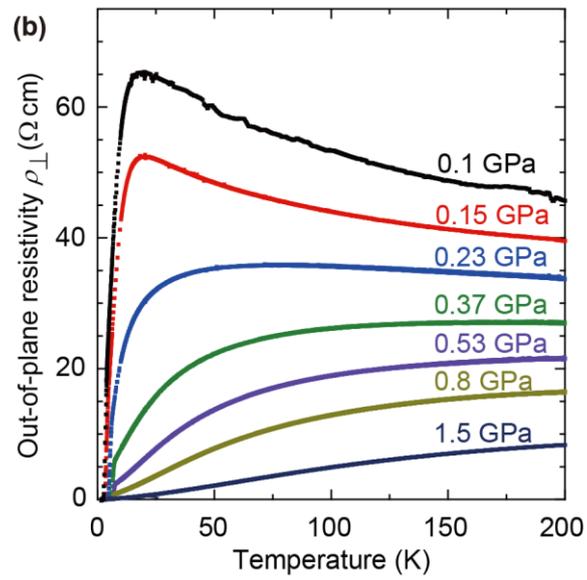

Fig. 3 Oike et al.

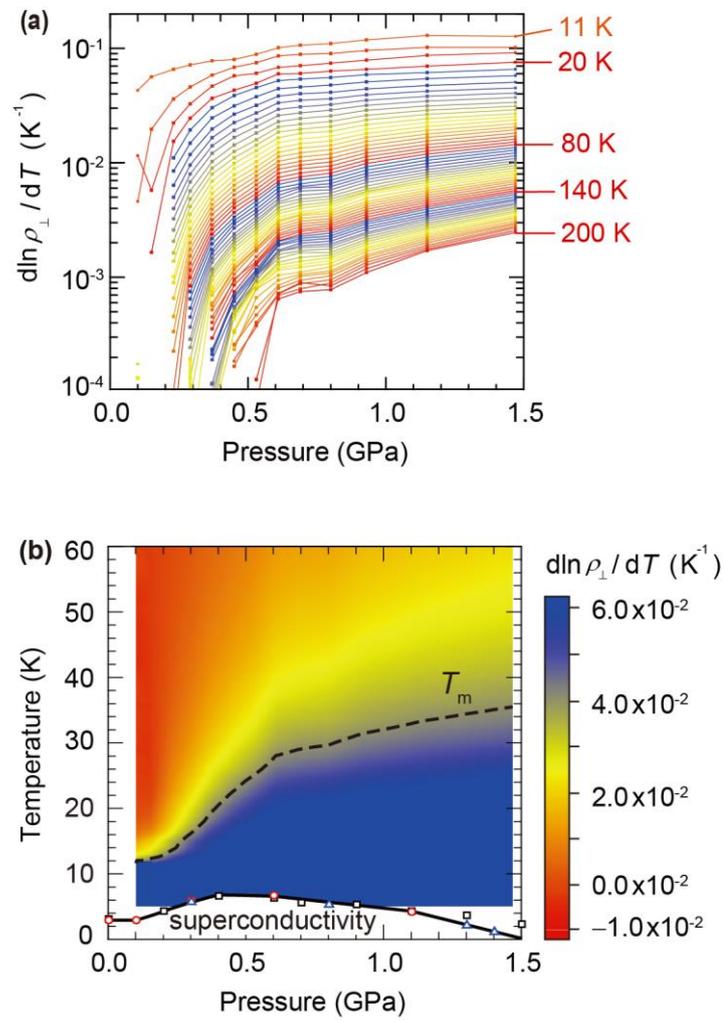

Fig. 4 Oike et al.

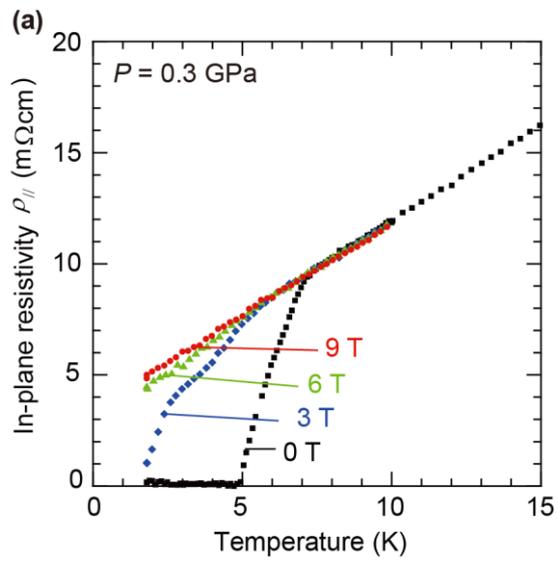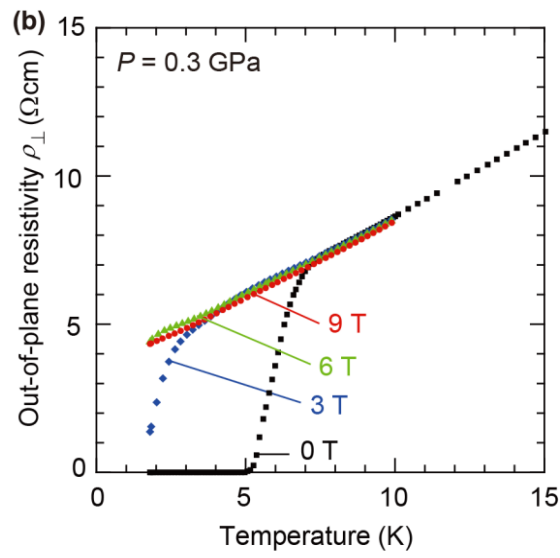

Fig. 5 Oike et al.